\documentclass[sigconf,nonacm]{acmart}
\AtBeginDocument{%
  }

\usepackage{microtype}
\usepackage{amsmath}

\usepackage{amssymb}
\usepackage{graphicx}
\usepackage{float}
\usepackage{subcaption}
\usepackage{colortbl}
\usepackage{multirow}
\usepackage{svg}

\definecolor{tblref}{HTML}{E8F5E9}
\definecolor{tblbest}{HTML}{E3F2FD}
\definecolor{tblbase}{HTML}{FFF3E0}
\definecolor{tblwarn}{HTML}{FFF9C4}
\definecolor{tblbad}{HTML}{FFEBEE}
\definecolor{tblhigh}{HTML}{E8EAF6}
\usepackage{enumitem}

\begin{document}

\title{Long-History User Transformers for Real-Time Ad Ranking}
\author{Viacheslav Ovchinnikov}
\affiliation{%
  \institution{Yandex}
  \city{Moscow}
  \country{Russia}}
\email{slavenchik@yandex-team.ru}

\author{Georgii Smirnov}
\affiliation{%
  \institution{Yandex}
  \city{Moscow}
  \country{Russia}}
\email{gsmirnov2@yandex-team.ru}

\author{Nikolai Savushkin}
\affiliation{%
  \institution{Yandex}
  \city{Moscow}
  \country{Russia}}
\email{penguin-diver@yandex-team.ru}

\author{Veronika Ivanova}
\affiliation{%
  \institution{Yandex, Applied AI Institute}
  \city{Moscow}
  \country{Russia}}
\email{veronika.ivanova88@yandex.ru}

\author{Maksim Kuzin}
\affiliation{%
  \institution{Yandex}
  \city{Moscow}
  \country{Russia}}
\email{kuzin@yandex-team.ru}

\renewcommand{\shortauthors}{Ovchinnikov et al.}
\begin{abstract}
Long interaction histories are among the most informative inputs for click-through rate (CTR) prediction, yet in online advertising they collide with a hard serving constraint: ads must be scored within a few hundred milliseconds to enter the auction, which rules out running a large sequence encoder at request time. We describe how a production advertising system resolves this conflict by decoupling history encoding from real-time inference. A high-capacity offline transformer asynchronously encodes the user's full cross-surface interaction history into a compact representation cached in a feature store, while a lightweight runtime model combines this cached representation with the user's most recent events and the request context at serving time. The offline encoder is pre-trained autoregressively on large-scale interaction logs with a dual objective---feedback prediction and next-item prediction---and the two-stage architecture is then fine-tuned for CTR prediction on the target advertising surface. Offline, the split design recovers 72--80\% of the quality of a full-history runtime transformer that would be too expensive to deploy, and the cached representation is robust enough to staleness to permit inexpensive refresh policies. In production A/B experiments, the system improves the primary ranking metric by \textbf{+2.77\%} in search advertising and \textbf{+2.1\%} on the Yandex Advertising Network, with revenue gains of \textbf{+2.26\%} and \textbf{+0.43\%} respectively---without increasing serving latency.
\end{abstract}


\maketitle
\section{Introduction}

User behavior in online advertising is inherently sequential, and click-through rate (CTR) models that exploit long interaction histories should outperform those that treat past behavior as a bag of features. Yet advertising rankers must finish within a few hundred milliseconds to participate in auctions, so fully processing long histories with transformers at serving time is too expensive. Recent transformer-based recommenders~\cite{zhai2024hstu,xiong2026llatte} show that long-history sequential models scale, but their deployment is easier in less latency-constrained settings. The challenge in ad ranking is therefore not only modeling long-range behavior, but doing so in a deployable form. We address this with a multi-stage transformer architecture: a large offline model asynchronously encodes the full user history into a fixed-size representation cached in a feature store, while a smaller runtime model consumes the freshest user events together with the cached embedding and current request features---preserving long-range signal while keeping online computation within the latency budget.

We pre-train the offline model autoregressively on cross-surface interaction logs following the ARGUS recipe~\cite{khrylchenko2026scaling} (feedback-prediction and next-item-prediction heads), then fine-tune the split architecture with a two-tower CTR objective on the target surface. Beyond offline quality, we study three operational questions: how much of the full-history ceiling the split recovers, how sensitive the cached representation is to staleness, and whether gains transfer across surfaces and hold in online A/B tests.

Our contributions are as follows:

\begin{enumerate}

\item We propose a multi-stage architecture that resolves the central deployment bottleneck for long-history transformer ranking in online advertising---full-history sequence modeling is beneficial for CTR prediction but too expensive for real-time serving---by moving heavy long-history computation offline and restricting online inference to a lightweight runtime model over recent events plus a cached offline representation.

\item We show that a split architecture with autoregressive pre-training recovers 72--80\% of the full-history quality ceiling across the two evaluated surfaces (Search Ads and YAN), while remaining deployable under strict ad-auction latency constraints.

\item We demonstrate in production that the resulting system delivers measurable gains under real latency constraints, improving the primary metric by +2.77\% on search advertising and +2.1\% on YAN.
\end{enumerate}


\section{Related Work}\label{sec:related}

Early neural CTR models: DeepFM~\cite{guo2017deepfm}, Wide~\&~Deep~\cite{cheng2016wide}, and DCN~V2~\cite{wang2021dcnv2} learn interactions among sparse user, item, and context features but treat history as aggregated statistics, losing temporal structure. We reuse DCN~V2 in the prediction head but feed it a sequential transformer encoding instead of a flat user vector. 
Sequential models address this by ordering user behavior: DIN~\cite{zhou2018deep} and DIEN~\cite{zhou2019deep} apply target-aware attention and GRU-based interest evolution, SIM~\cite{pi2020search} retrieves relevant subsequences from long histories, and production systems such as TransAct~\cite{xia2023transact} and PinnerFormer~\cite{pancha2022pinnerformer} run small transformers over short windows (up to a few hundred events) to meet latency budgets. Our split design extends the effective history to 8\,192 events by moving long-range computation offline.


Recent work shows that recommender transformers scale like language models: HSTU~\cite{zhai2024hstu} reformulates recommendation as generative sequential transduction over 8\,000-interaction histories; LLaTTE~\cite{xiong2026llatte} establishes scaling laws for multi-stage sequence models in ads; Khrylchenko et~al.~\cite{khrylchenko2026scaling} scale recommender transformers via dual-objective autoregressive pre-training (feedback prediction and next-item prediction), which we adopt and extend to cross-surface advertising data; PinRec~\cite{badrinath2025pinrec} extends the generative paradigm to retrieval with multi-token prediction.


Our work combines these ideas in online advertising: we split computation offline/online like LLaTTE and TransAct and pre-train a causal transformer with a dual objective like HSTU and Khrylchenko et~al., contributing a unified cross-surface timeline with heterogeneous tokenization and a deployment-oriented split design that recovers roughly 80\% of full-history quality under production latency.

\section{Methodology}

We study click-through rate (CTR) prediction for online advertising. Given a user $u$ and a candidate banner $b$, the model predicts the probability of a click:

\begin{equation}\label{eq:ctr}
  \hat{y} = f(S_u,\, b;\, \boldsymbol{\theta}) \in [0, 1],
\end{equation}
where $S_u$ is the user's interaction history (Section~\ref{sec:user-history}), $b$ is the candidate banner, and $\boldsymbol{\theta}$ are the learned parameters of the model. Internally, $f$ encodes $S_u$ into a user-side embedding $\mathbf{z}_{\text{user}}$ and $b$ into a candidate embedding $\mathbf{z}_{\text{b}}$; the final score is then their inner product (Eq.~\ref{eq:fusion}).

To resolve the latency--quality tension described in the introduction, we adopt a two-stage design: a large offline model asynchronously extracts long-range behavioral signals, while a lightweight runtime model focuses on fresh events available at request time.

\subsection{History Construction}\label{sec:user-history}

From logs, we build a unified interaction history $S_u = \{a_1, a_2, \dots, a_T\}$ spanning both organic behavior (search clicks and user search queries) and advertising surfaces (Yandex Advertising Network, Product Gallery, Search Ads).
Here, Yandex Advertising Network (YAN) is Yandex's external advertising network, covering programmatic ad placements across partner websites, mobile applications, and other partner surfaces (e.g., third-party messaging clients). Throughout the paper, ``YAN'' refers to the full external advertising network and is the unit on which we report online A/B metrics, since we want to capture the integral business impact across the network. All offline experiments and the production rollout, however, were carried out on a single web-only slice of that network, which we refer to as \emph{YAN web inventory}. Product Gallery is a carousel of product cards from multiple merchants shown on the search results page. Search Ads are sponsored website listings shown above the organic search results. Merging these channels into a single timeline gives the model a broader view of user intent than any single surface alone.

Each action $a_t$ records a timestamp, surface or interaction type, banner identifier when applicable, associated text embeddings, and a click/non-click label. Text embeddings are produced by a shared embedding service that maps a bag-of-words representation of the item's text (e.g., banner title and description, search query) to a dense vector; the exact dimensionality is reported together with the rest of the model configuration in Section~\ref{sec:config}.

\paragraph{Awareness regimes.}
Before describing the tokenization, it is useful to fix terminology for how much information about the current request is exposed to the user encoder. We distinguish three regimes:

\begin{itemize}
\item \textbf{NoAware}: the encoder sees only the user's history; the request context (page, surface, device) and the candidate banner are kept out of the sequence. The history hidden state is consumed by the prediction head as-is. Because the user representation depends only on past behavior, it can be precomputed entirely offline and refreshed asynchronously, so request-time scoring reduces to an inner product with the candidate embedding and there is no transformer forward pass on the critical path. This is the cheapest setup at serving time and the strictest probe of how much signal the history alone carries.
\item \textbf{ContextAware}: the encoder additionally sees the current request context, but not the specific candidate banner. A single forward pass per request yields a user-side representation that can be scored against many candidates.
\item \textbf{TargetAware}: each candidate banner is injected into the sequence and the encoder is re-run per candidate. This is the most informative regime but the most expensive at serving time --- the runtime cost scales with the candidate set size.
\end{itemize}

\begin{figure*}[t]
  \centering
  \includegraphics[width=2.0\columnwidth]{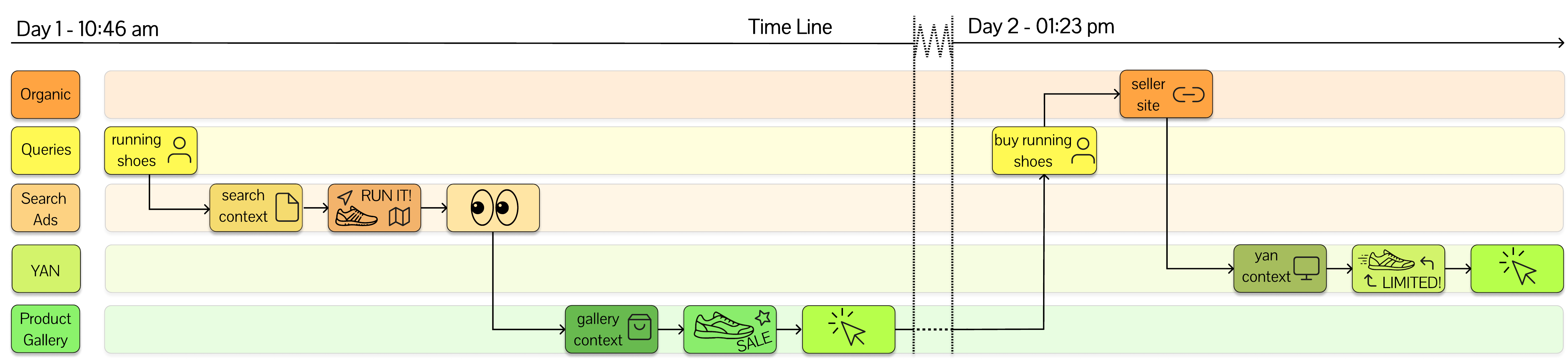}
  \caption{A two-day timeline of cross-surface user interactions. Rows
  correspond to channels (organic, queries, Search Ads, YAN, Product Gallery).
  Blocks denote events in temporal order. Organic interactions use a single token, while advertising surfaces carry three fields (\textit{context}, \textit{banner}, \textit{action}); for transformer input the \textit{banner} and \textit{action} embeddings are summed into a single vector, yielding two input tokens per advertising event.}
  \label{fig:timeline}
\end{figure*}

Our deployable design uses ContextAware: it turned out to be the best quality--cost trade-off for ad ranking, where the candidate set per request is large enough that TargetAware becomes prohibitive, while NoAware leaves a substantial amount of context-conditioned signal on the table.
\paragraph{Sequence tokenization.}
Figure~\ref{fig:timeline} illustrates the unified sequence representation and
the tokenization scheme applied prior to feeding the sequence into the
transformer. Organic search clicks and user queries are each encoded as a single token.
Advertising interactions across YAN, Product Gallery, and Search Ads are
described by three fields — \textit{context} (request-side features: page,
device, surface), \textit{banner} (creative identity), and \textit{action}
(user feedback) — which together separate request context, creative
identity, and user feedback within a single autoregressive sequence.

Because the deployed model is ContextAware, it is the \textit{context} field that the encoder must see explicitly: the candidate banner is never placed in the sequence at request time, so there is no need to keep \textit{banner} and \textit{action} as separate input tokens. We therefore sum the \textit{banner} and \textit{action} embeddings into a single input vector, so each
advertising event contributes two input tokens: a \textit{context} token
followed by a combined \textit{banner+action} token. This reduces the
advertising portion of the sequence by a third relative to a fully
unrolled three-token representation and proportionally lowers training
and inference cost on advertising-heavy users.

\subsection{Pre-training}\label{sec:pretraining}

The goal of pre-training is to learn a shared sequential representation that can initialize both encoders used at serving time. We therefore pre-train a single large transformer with a causal attention mask and ALiBi positional encoding~\cite{press2022alibi} on unified cross-surface histories collected over one year, following the general paradigm of generative recommenders~\cite{zhai2024hstu} and the dual-objective autoregressive pre-training recipe of Khrylchenko et~al.~\cite{khrylchenko2026scaling}. In our setting, the mask is applied over the merged timeline of user actions across all surfaces and interaction types. The mask operates at the token level: each position attends to all preceding tokens regardless of which surface they came from or how many tokens a given event contributes (a single token for organic clicks and queries, two tokens for advertising events --- a \textit{context} token followed by a combined \textit{banner+action} token formed by summing the banner and action embeddings). The model therefore sees one unified token stream and makes no architectural distinction between event types in the attention pattern itself.

The model is trained in a multi-task multi-surface regime. Each advertising surface (Product Gallery, Search Ads, YAN) carries both a CTR head (Eq.~\ref{eq:bce}) and contributes to the retrieval objective (Eq.~\ref{eq:retrieval}). Organic clicks and search queries do not have CTR labels, but they are \emph{not} passive context during pre-training: at each organic position the model is asked to predict the next organic click or query from the preceding hidden state under the same sampled-softmax retrieval loss. Every token in the sequence therefore contributes to some pre-training objective.

\paragraph{Click prediction.} We use a standard binary cross-entropy loss over observed click labels. For each example $i$, let $y_i \in \{0,1\}$ denote the observed click label (1 if the user clicked, 0 otherwise), and let $\hat{y}_i$ denote the predicted click probability:
\begin{equation}\label{eq:bce}
  \mathcal{L}_{\text{CTR}} = -\frac{1}{N}\sum_{i=1}^{N}\bigl[y_i \log \hat{y}_i + (1-y_i)\log(1-\hat{y}_i)\bigr].
\end{equation}
During pre-training, the CTR objective uses a single joint head rather than the two-tower decomposition used at fine-tuning (Section~\ref{sec:architecture}). At each advertising position, this head concatenates the transformer hidden state $\mathbf{h}_t$ on the user side with the banner-side input embedding $\mathbf{e}_b$ --- the same vector that feeds the \textit{banner} field of the corresponding event token --- and a linear layer outputs the logit. The two-tower vectors $\mathbf{z}_{\text{user}}, \mathbf{z}_b$ from Eq.~\ref{eq:fusion} are not yet defined at this stage: they appear only when the two CTR towers are introduced for fine-tuning.

\paragraph{Next-event retrieval.} The pre-trained backbone is designed to initialize both the CTR ranker studied here and separate candidate-generation models on individual surfaces; the latter require an inner-product embedding geometry that a purely BCE-trained representation does not enforce. We therefore add an autoregressive retrieval objective alongside the CTR head: at each position the model predicts the user's next engagement event of the same type (next clicked banner for advertising positions, next organic click or query for organic positions). For each event type (modality) $m$, the inner-product geometry is realized by a pair of dedicated projection heads $g_{\text{user}}^{(m)}, g_{\text{tgt}}^{(m)}$ that map the transformer hidden state $\mathbf{h}_t$ and the target input embedding $\mathbf{e}_x$ into a shared modality-specific scoring space:
\begin{equation*}
  \mathbf{z}_t^{(m)} = g_{\text{user}}^{(m)}(\mathbf{h}_t), \qquad \mathbf{z}_x^{(m)} = g_{\text{tgt}}^{(m)}(\mathbf{e}_x).
\end{equation*}
These per-modality heads are distinct from the two-tower CTR heads used at fine-tuning (Eq.~\ref{eq:fusion}); they are trained only during pre-training and are not used at serving time. Training uses sampled softmax with the $\log Q$ correction~\cite{yi2019sampling} to account for sampling bias:

\begin{equation}\label{eq:retrieval}
  \mathcal{L}_{\text{ret}}^{(m)} = -\log \frac{\exp\bigl(\langle \mathbf{z}_t^{(m)}, \mathbf{z}_{x^+}^{(m)} \rangle - \log Q(x^+)\bigr)}
  {\displaystyle\sum_{x' \in \mathcal{X}_t} \exp\bigl(\langle \mathbf{z}_t^{(m)}, \mathbf{z}_{x'}^{(m)} \rangle - \log Q(x')\bigr)},
\end{equation}

where $\langle\cdot,\cdot\rangle$ is the inner product, $\mathbf{e}_x$ is the input embedding of a candidate target (banner for advertising positions, organic item or query for organic positions), $x^+$ is the positive next target of the appropriate type, $\mathcal{X}_t$ is the set of sampled negatives drawn from the same type vocabulary together with $x^+$, and $Q(x)$ is the per-type sampling distribution. We estimate $Q$ online using the streaming frequency estimator with hash-based counting~\cite{yi2019sampling}. In practice, $\mathcal{X}_t$ is constructed from in-batch negatives---candidate targets of the same type appearing in other examples of the per-device batch are reused as negatives---at no additional sampling cost, so the size of $\mathcal{X}_t$ varies by event type and reflects how frequent that type is in a batch.

The combined pre-training loss across all surfaces is
\begin{equation}\label{eq:pretrain}
  \mathcal{L}_{\text{pre}} = \sum_{m \in \mathcal{M}_{\text{ads}}} \mathcal{L}_{\text{CTR}}^{(m)} \;+\; \lambda \sum_{m \in \mathcal{M}_{\text{all}}} \mathcal{L}_{\text{ret}}^{(m)},
\end{equation}
where $\mathcal{M}_{\text{ads}} = \{\text{Product Gallery, Search Ads, YAN}\}$ are the surfaces with click labels and $\mathcal{M}_{\text{all}} = \mathcal{M}_{\text{ads}} \cup \{\text{organic clicks, queries}\}$ are all event types that contribute to retrieval.
The CTR objective teaches the model to predict banner clickability, while the retrieval objective encourages the hidden state to preserve information about the item with which the user is likely to engage next. 

This dual-objective formulation mirrors the feedback prediction and next-item prediction tasks identified by Khrylchenko et~al.~\cite{khrylchenko2026scaling} as an effective pre-training recipe for recommender transformers. We fix $\lambda = 1$; the two loss terms are on different absolute scales in practice, but we do not tune $\lambda$ because each pre-training run is computationally expensive.

\subsection{Multi-Stage Architecture and Fine-tuning}\label{sec:architecture}

The two-stage architecture is the main mechanism used to resolve the latency--quality trade-off. Whereas Khrylchenko et~al.~\cite{khrylchenko2026scaling} focus on scaling a single large autoregressive recommender model, our setting requires a deployable split design in which long-history computation is performed asynchronously offline and only a short suffix is processed at request time. The architecture is designed to satisfy three objectives simultaneously: (i) capture as much long-term behavioral signal as possible, (ii) remain sensitive to fresh user intent, and (iii) keep request-time computation within the production latency budget.

\begin{figure*}[t]
  \centering
  \includegraphics[width=2.0\columnwidth]{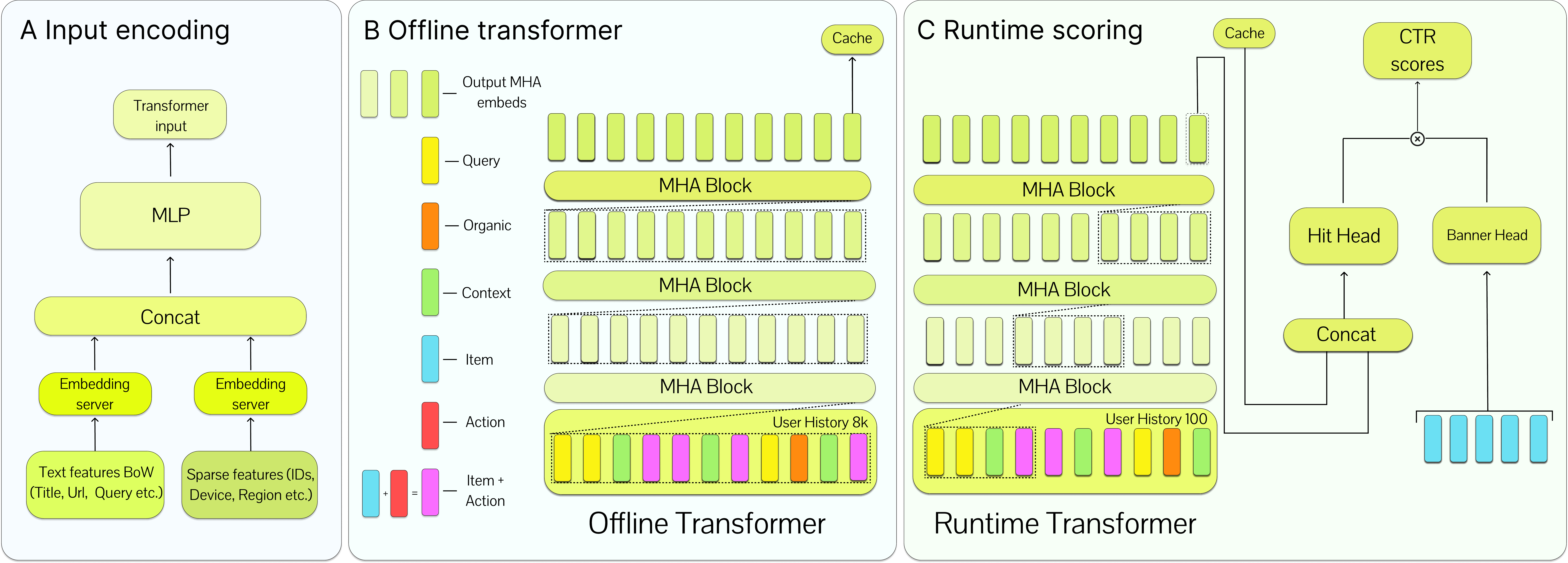}
  \caption{Multi-stage architecture overview. (A) Input encoding maps text and sparse features into transformer tokens. (B) The offline transformer processes long user histories asynchronously and caches fixed-size representations. (C) At request time, the runtime transformer encodes recent actions, combines its output with the cached offline representation in the user head, and scores it against the banner head to produce CTR scores.}
  \label{fig:architecture}
\end{figure*}

After pre-training, we instantiate two separate transformers for fine-tuning on a target advertising surface from the same shared $L_{\text{off}}$-layer checkpoint (Figure~\ref{fig:architecture}). The offline model retains the full depth, whereas the runtime model is an $L_{\text{rt}}$-layer network ($L_{\text{rt}} \le L_{\text{off}}$) initialized from the first $L_{\text{rt}}$ layers of that checkpoint. This initialization scheme was chosen empirically: copying the lower layers of the pre-trained stack consistently yielded lower CTR fine-tuning loss than initializing the runtime encoder from scratch. The concrete values of $L_{\text{off}}$ and $L_{\text{rt}}$ used in our deployment are reported in Section~\ref{sec:config}. The two encoders also differ in their handling of position: the offline transformer inherits ALiBi from the pre-trained backbone (Section~\ref{sec:pretraining}), while the runtime transformer uses no explicit positional encoding---its input is short and the receptive-field window already imposes strict locality, so we found a positional signal unnecessary in practice. The two encoders are then fine-tuned jointly with the prediction head for the downstream CTR task.

\paragraph{Offline model.} The offline transformer asynchronously encodes the full user history $S_u$ into a single fixed-size vector. The encoder maps the history to hidden states and we take the last one as $\mathbf{z}_{\text{off}} \in \mathbb{R}^{d}$. This vector preserves long-range information that would be too expensive to recompute online.

\emph{At training time}, we deliberately simulate the inevitable staleness of the production cache. When constructing a training example for impression time $t$, we feed the offline encoder only those events of the user's history that occurred more than two days before $t$; the resulting $\mathbf{z}_{\text{off}}$ is then concatenated with the runtime encoder's output and passed to the prediction head. This two-day cutoff is dictated by the offline pipeline cycle in production---roughly one day to ingest and deduplicate raw logs, plus another day to merge user histories, run batch inference, and propagate the resulting vectors to the feature store---so this construction makes the training distribution of $\mathbf{z}_{\text{off}}$ match the lag that the model will actually see at serving time.

\emph{At serving time}, the offline pipeline runs asynchronously, writes $\mathbf{z}_{\text{off}}$ to a feature store, and the ranker reads whichever vector is currently cached for the user---without applying any artificial cutoff. The vector is refreshed event-driven when the user clicks an ad, and a full batch recomputation runs once per week for the long tail of less active users; the full refresh policy is described in Section~\ref{sec:exp-staleness}.

\paragraph{Runtime model.} The runtime transformer encodes the most recent suffix of the user's history at request time and captures short-horizon intent that may change between the last offline refresh and the current request. The suffix is defined by a fixed event count, not by a hard cutoff: the runtime model always consumes the latest available events regardless of whether some of them predate the offline cutoff. For active users this suffix is dominated by events after the cutoff; for users with no activity in the last two days, it falls entirely before the cutoff and overlaps with the prefix already summarized by the offline encoder, leaving no gap in the user's representation. The runtime encoder's receptive field is deliberately constrained; the windowed-attention mechanism and the choice of window size are detailed in Section~\ref{sec:receptive}. The output of the runtime transformer over this suffix is taken as $\mathbf{z}_{\text{rt}} \in \mathbb{R}^{d}$ in the same way as for the offline encoder (final-position hidden state).


\paragraph{Fusion and prediction.} The two encoders are fine-tuned end-to-end together with the prediction head using gradients from the CTR objective (training details in Section~\ref{sec:exp-training}). During fine-tuning we switch from the pre-training's joint head (which fuses user-side and candidate-side inputs into a single logit) to a two-tower decomposition: a user tower and a candidate (banner) tower whose outputs are scored by inner product. The two-tower form is needed for deployment---candidate-side embeddings can be precomputed and cached---and lets the same backbone power downstream candidate-generation models. The user tower $g(\cdot)$ takes the two transformer outputs $\mathbf{z}_{\text{off}}, \mathbf{z}_{\text{rt}} \in \mathbb{R}^{d}$, concatenates them, and projects the resulting $2d$-dimensional vector down to a scoring vector $\mathbf{z}_{\text{user}} = g([\mathbf{z}_{\text{off}};\,\mathbf{z}_{\text{rt}}]) \in \mathbb{R}^{d_{\text{tower}}}$ via a DCNv2~\cite{wang2021dcnv2} block followed by an MLP. The candidate tower has the same DCNv2 + MLP structure but takes the banner-side item embedding (the same vector that feeds the \textit{banner} token of the transformer) as input and produces $\mathbf{z}_{\text{b}} \in \mathbb{R}^{d_{\text{tower}}}$ in the same scoring space. The value of $d_{\text{tower}}$ is reported in Section~\ref{sec:config}.

\begin{equation}\label{eq:fusion}
  \hat{y} = \sigma\!\bigl(\mathbf{z}_{\text{user}}^\top \mathbf{z}_{\text{b}}\bigr),
  \quad \mathbf{z}_{\text{user}} = g([\mathbf{z}_{\text{off}};\, \mathbf{z}_{\text{rt}}]).
\end{equation}
The cached offline representation and the runtime representation play complementary roles: the former summarizes long-term behavior, while the latter injects fresh request-time evidence.

\subsection{Model Configuration}\label{sec:config}

Table~\ref{tab:config} summarizes the architectural hyperparameters of both stages and the prediction head. The chosen configuration reflects the latency constraints of the runtime stage and the capacity requirements of the offline encoder.

\begin{table}[t]
  \centering
  \caption{Encoder configuration. The two transformers share the same hidden dimension but differ in depth and sequence scope. The bottom row reports $d_{\text{tower}}$, the output dimension of the shared two-tower prediction head and therefore a single value for the model as a whole rather than a per-encoder setting.}
  \label{tab:config}
  \begin{tabular}{lcc}
      \hline
      & Offline & Runtime \\
      \hline
      Input embedding dim & 64    & 64 \\
      Hidden dim $d$      & 1024  & 1024 \\
      Positional encoding & ALiBi & none \\
      Layers ($L_{\text{off}}$, $L_{\text{rt}}$) & 10 & 5 \\
      Window size $w$     & 8192   & 20 \\
      Receptive field $R$ & 8\,192 & 100 \\
      Sequence at serving & full history & recent events \\
      \hline
      Two-tower output dim $d_{\text{tower}}$ & \multicolumn{2}{c}{128} \\
      \hline
    \end{tabular}
\end{table}

The full model contains approximately 200\,M trainable parameters.

\subsection{Receptive Field Design}\label{sec:receptive}

A key design constraint of the runtime encoder is its limited receptive field. The runtime transformer must cover enough of the user's most recent events to capture short-term intent while remaining within a strict latency budget. We therefore constrain its receptive field by design: for a runtime encoder with $L_{\text{rt}}$ layers and per-layer window size $w$, the total receptive field is $R = L_{\text{rt}} \times w$---each output position attends only to the latest $R$ positions (the most recent events in the user's history), independently of how many events the user has accumulated. In our deployment we set $L_{\text{rt}} = 5$, $w = 20$ ($R = 100$): the production pipeline can feed up to 200 events at request time, but the ablation in Section~\ref{sec:exp-search} shows that doubling the receptive field yields only a moderate gain (+12.3\% vs.\ +11.7\% for the split configuration), so we use $R = 100$ as a coverage--cost compromise.

\subsection{Computational Complexity}\label{sec:complexity}

Two factors determine the cost of attention in our system: sequence length and per-layer attention pattern. Vanilla causal self-attention scales as $\mathcal{O}(T^2)$ per user in sequence length $T$. To bound this cost on long histories and to make the runtime latency budget enforceable, the runtime model uses \emph{windowed} causal attention: at each layer every position attends only to the preceding $w$ tokens, yielding $\mathcal{O}(T \cdot w)$ cost per layer. Stacking $L_{\text{rt}}$ such layers produces a receptive field of $R = L_{\text{rt}} \cdot w$ tokens, so depth and window width can be traded off explicitly against the serving budget of a particular surface.

The offline model, by contrast, uses full causal attention over the entire 8\,192-token prefix. Its $\mathcal{O}(T^2)$ cost is acceptable because the offline pipeline runs asynchronously on dedicated GPUs once per refresh, not once per request. Across $N$ users the total offline cost is $\mathcal{O}(N \cdot T^2)$: quadratic in sequence length, linear in the number of users, which keeps the approach practical even at the scale of hundreds of millions of users.

\section{Experiments}

Our experimental evaluation is organized around six research questions:
\begin{itemize}
    \item (RQ1) How effectively does the split architecture recover the quality of a full-history transformer that cannot be deployed in production?
    \item (RQ2) Given the split architecture, how sensitive is quality to training recipe choices such as data composition,  fine-tuning and position debiasing?
    \item (RQ3) How sensitive is the cached offline representation to embedding and model staleness?
    \item (RQ4) Do the offline findings replicate on a second advertising surface?
    \item (RQ5) Does the magnitude of the gain track the amount of available user history, as the central premise of the paper predicts?
    \item (RQ6) Does the offline quality improvement translate to measurable online gains, and does the split design preserve production latency?
\end{itemize}

\subsection{Evaluation Setup}

We evaluate the transformer-based sequence model by injecting its prediction as an additional feature into the production CTR model---a CatBoost~\cite{prokhorenkova2018catboost} ranker with on the order of $10^3$ hand-engineered and learned features. This protocol isolates the incremental value of the sequence representation: any improvement can be attributed to information captured by our model, rather than to replacing the existing production feature set.

\paragraph{Metric.} We adopt a cross-entropy-based offline metric in the spirit of Normalized Entropy~\cite{he2014practical}, rather than AUC. In production ranking systems, predicted CTRs feed directly into auction mechanics, budget pacing, and revenue optimization, so calibration matters as much as ordering; unlike AUC, log-loss penalizes miscalibrated predictions and is therefore a closer offline proxy for online business impact~\cite{he2014practical}. Concretely, we report a normalized log-loss difference between the production CTR ranker with and without our transformer-based feature:
\begin{equation}\label{eq:nll}
  \Delta_{\text{NLL}} = \frac{\mathcal{L}_{\text{prod}} - \mathcal{L}}{C} \cdot 100\%,
\end{equation}
where $\mathcal{L}_{\text{prod}}$ is the log-loss of the production CatBoost ranker on its existing feature set, $\mathcal{L}$ is the log-loss of the same ranker after our model's prediction is added as an additional feature, and $C$ is the total number of clicks in the evaluation set. A positive $\Delta_{\text{NLL}}$ thus quantifies the incremental contribution of the transformer feature to the production system; higher is better.

\paragraph{Corpus composition.} In the pre-training corpus, search queries dominate the event stream ($\sim$40\%) and organic clicks contribute the next-largest share ($\sim$30\%); among advertising surfaces, Search Ads and YAN are roughly comparable in scale ($\sim$13--14\%) while Product Gallery is the rarest ($\sim$3\%). These per-event shares reflect each surface's contribution to the pre-training loss (Eq.~\ref{eq:pretrain}).


\paragraph{Choice of primary testbed.} Most architectural and training ablations were conducted on desktop Search Ads. Its production CatBoost model is trained on a relatively small, position-bias-cleaned traffic slice, which makes retraining and evaluation substantially faster than on other surfaces. We found that nearly all conclusions transferred reliably to Yandex Advertising Network web inventory, i.e., ads served through YAN on partner websites rather than generic web traffic. We therefore present the full ablation study on Search Ads and use YAN web inventory as the main transfer setting (Section~\ref{sec:exp-web}).
\subsection{Training Setup}\label{sec:exp-training}

\paragraph{Optimizer and schedule.} Both pre-training and fine-tuning use the Adam optimizer with no warmup, no weight decay, and dropout disabled. The learning rate starts at $2 \times 10^{-4}$ and is decayed by a factor of $0.5$ three times during training, at intervals of approximately one-fifth of an epoch.

\paragraph{Pre-training.} Pre-training is run for 1 epoch over the full corpus described in Section~\ref{sec:user-history} (approximately one billion events collected over one year). Training is performed on 64 NVIDIA A800 GPUs with batches packed to $\sim$50\,000 tokens \emph{per GPU}, and takes about two days of wall-clock time.

\paragraph{Fine-tuning.} Fine-tuning is run for 3 epochs on the target advertising surface, with end-to-end gradients through both encoders and the prediction head. Together with the preceding pre-training epoch, the total training budget is $1 + 3$ epochs.

\subsection{RQ1: How effectively does the split architecture recover the quality of a full-history transformer that cannot be deployed in production?}\label{sec:exp-search}

Table~\ref{tab:search-ablation} summarizes the main architectural ablation on desktop search. We vary two factors: the capacity of the runtime transformer (full vs.\ half-size) and the number of events in its receptive field (100 vs.\ 200). The \emph{Full model (8k)} row provides an upper bound: a single large transformer consumes the full 8\,192-event history at inference time. Although this configuration is not deployable in production, it serves as a useful reference for measuring how much of the full-history gain is recovered by the split design.

\begin{table}[!ht]
  \centering
  \caption{Architecture ablation on desktop search advertising. $\Delta_{\text{NLL}}$ (\%) relative to the production baseline.}
  \label{tab:search-ablation}
  \begin{tabular}{llcl}
    \hline
    Configuration & Runtime events & $\Delta_{\text{NLL}}$ (\%) & \\
    \hline
    \multicolumn{4}{l}{\textit{Single-stage (online only)}} \\
    \rowcolor{tblref}
    Full model (8k)       & 8\,000 & +16.2 & ceiling \\
    Full transformer      & 200    & +7.29 & \\
    Full transformer      & 100    & +6.30 & \\
    Half-size transformer & 200    & +7.01 & \\
    Half-size transformer & 100    & +5.75 & \\
    \hline
    \multicolumn{4}{l}{\textit{Split scheme (offline 8k + online)}} \\
    \rowcolor{tblbest}
    Half-size + offline   & 200    & +12.3 & best \\
    Half-size + offline   & 100    & +11.7 & \\
    \hline
  \end{tabular}
\end{table}

Several patterns emerge. Increasing the runtime receptive field from 100 to 200 events consistently improves quality, although the gain is moderate in magnitude. Reducing model size has only a limited effect when the receptive field is 200 events (+7.01\% vs.\ +7.29\%), suggesting that, at this scale, sequence coverage matters more than runtime model width. Most importantly, the split scheme recovers most of the full-history benefit. With 200 runtime events and a half-size runtime transformer, the best deployable configuration reaches $+12.3\%$, closing $76\%$ of the gap to the impractical 8k single-stage upper bound ($+16.2\%$). The shorter $R=100$ configuration loses only $0.6$\,pp to this best variant ($+11.7\%$, i.e.\ $72\%$ retention) and is the one used in production for a tighter runtime attention budget (Section~\ref{sec:exp-web}). Relative to the corresponding runtime-only setup, the cached offline representation contributes an additional +5.3\,pp (+12.3\% vs.\ +7.01\%), indicating that it captures long-range behavioral signal that cannot be reconstructed from the short runtime window alone.

\subsection{RQ2: Given the split architecture, how sensitive is quality to training recipe choices such as data composition,  fine-tuning and position debiasing?}\label{sec:exp-recipe}

Architecture alone is not sufficient to achieve the best quality. Table~\ref{tab:recipe-ablation} traces the gains on desktop Search Ads CTR as we refine the training recipe. The starting point is a multi-task model trained jointly on three surfaces: Product Gallery, Search Ads, and YAN.

\begin{table}[!ht]
  \centering
  \caption{Training recipe ablation on desktop search. Each row adds one modification on top of the previous one.}
  \label{tab:recipe-ablation}
  \begin{tabular}{lcl}
    \hline
    Training configuration & $\Delta_{\text{NLL}}$ (\%) & \\
    \hline
    \rowcolor{tblbase}
    Multi-task & +7.5 & baseline \\
    \quad + single-modality fine-tuning & +8.5 & \\
    \quad + organic click events        & +11.0 & \\
    \rowcolor{tblref}
    \quad + position debiasing \& LR    & +15.0 & reference \\
    \hline
  \end{tabular}
\end{table}

\paragraph{Single-modality fine-tuning.} The pre-trained model is first optimized with a joint loss across Product Gallery, Search Ads, and YAN. Fine-tuning exclusively on the target surface (Search Ads) adds +1.0\,pp (+7.5\% $\to$ +8.5\%), suggesting that the shared objective introduces some degree of gradient interference and weakens surface-specific adaptation at the final stage.

\paragraph{Incorporating organic clicks.} Earlier iterations of the recipe used only advertising events; this row adds organic (non-advertising) search clicks to the user history and yields a further +2.5\,pp improvement (+8.5\% $\to$ +11.0\%). Organic clicks are both abundant and highly informative: when a user clicks an organic result, that action often reveals intent that is predictive of subsequent ad engagement.
This observation is consistent with LLaTTE, which reports that enriching the input token distribution improves scaling behavior~\cite{xiong2026llatte}.

\paragraph{Position debiasing and learning rate.} The true display position is provided as a feature to the transformer for every historical event, both during training and at inference. The difference between training and inference is only in how the position of the \emph{candidate being scored} is handled: during training it is set to its true value, but at inference time it is fixed to~1, so the model is asked to predict CTR as if the candidate were shown in the top position regardless of where the ad would actually appear. In effect, this reduces the confounding influence of position on click probability and encourages the model to predict intrinsic relevance rather than position-inflated CTR \cite{guo2019pal}. Combined with a higher learning rate, this change produces the final +4.0\,pp gain (+11.0\% $\to$ +15.0\%), the largest improvement among the individual recipe components.

Taken together, these results show that data composition and optimization choices are at least as important as the model architecture itself for realizing the benefit of long-history representations.

\subsection{RQ3: How sensitive is the cached offline representation to embedding and model staleness?}\label{sec:exp-staleness}

A practical concern in the two-stage design is freshness. Because the offline representation is computed asynchronously rather than at request time, it inevitably lags behind the user's most recent behavior. If its quality degrades rapidly, the system would require frequent recomputation, increasing infrastructure cost and undermining the main advantage of the split architecture.

\begin{figure}[t]
  \centering
  \includegraphics[width=\columnwidth]{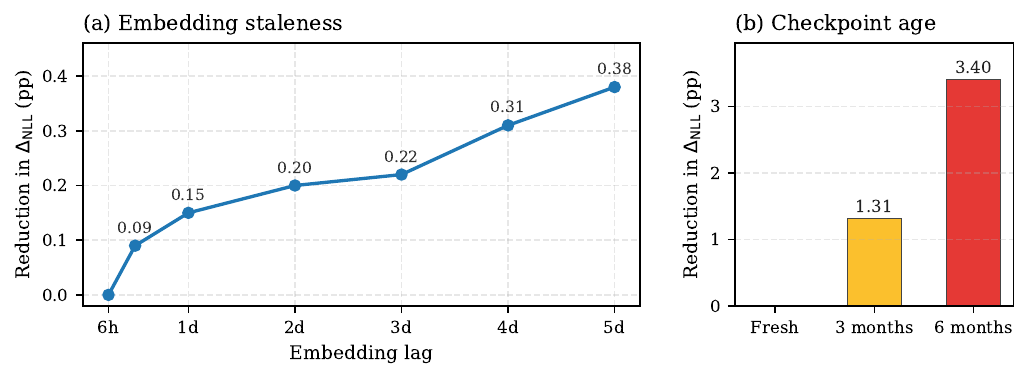}
  \caption{Operational staleness of the offline component. Each value is the reduction in $\Delta_{\text{NLL}}$ (percentage points) relative to the freshest configuration. (a) Cached embedding lag: degradation accumulates gradually over days, with only $0.38$\,pp lost after 5 days. (b) Model checkpoint age: a 6-month-old checkpoint loses $3.40$\,pp, an order of magnitude more than the worst embedding-lag setting. Regular model retraining is therefore much more important than aggressive embedding refresh.}
  \label{fig:staleness}
\end{figure}

To quantify this effect, we evaluate the offline model on a held-out desktop Search Ads slice while varying the embedding lag (Figure~\ref{fig:staleness}). We also study a related but distinct form of staleness: the effect of using an outdated offline model checkpoint. Both results are reported as reductions in gain relative to the freshest setting.


The embedding results indicate substantial robustness: even after a 5-day lag, the total reduction in gain is only 0.38, and the degradation accumulates gradually rather than collapsing at any particular lag. Model staleness, in contrast, is much more damaging: a checkpoint that is 6 months old loses 3.40 points of gain. In operational terms, regular retraining is therefore much more important than aggressive embedding refresh.

We also evaluate the \emph{NoAware} setup introduced in Section~\ref{sec:user-history}: the offline model encodes user history with a 2-day cutoff, the last hidden vector is passed through the prediction head, and the result is dotted with the banner embedding. No runtime model and no request context (device, page, surface) is provided, so the setup isolates the predictive signal carried by ad interaction history alone. Even under these constraints, it achieves +2.81\% $\Delta_{\text{NLL}}$ on the held-out search slice. This is notable because search advertising is strongly query-conditioned: the same user may be shown very different ads for different queries, so one might expect the current context to dominate. The fact that history alone still yields a substantial gain suggests that long-term ad engagement patterns contain strong predictive signal even when the current query is unavailable.

\paragraph{Production refresh policy.} Guided by these observations, we adopt a hybrid refresh strategy in production. Offline embeddings are recomputed \emph{on click}: whenever a user clicks an ad, the offline model re-encodes the updated history and writes a fresh embedding to the feature store. This event-driven policy ensures that the most engaged users are refreshed promptly. For the remaining users, a full batch recomputation runs once per week to prevent gradual staleness accumulation. This design reflects a deliberate trade-off: refreshing on every event would maximize freshness but at substantially higher offline compute cost. Since embedding quality degrades slowly, the click-triggered policy captures most of the benefit at a much lower cost.

\subsection{RQ4: Do the offline findings replicate on a second advertising surface?}\label{sec:exp-web}

Table~\ref{tab:web-ablation} reports transfer results on \emph{YAN web inventory}. The online A/B test in Section~\ref{sec:exp-ab} subsequently evaluated the same model on the full YAN network. Absolute gains are smaller than on desktop Search Ads, and the two surfaces differ in a structural way that explains this gap: Search Ads are shown in response to an explicit user query, which is itself a strong predictor of click intent and pushes the baseline CTR substantially higher. YAN ads, by contrast, are placed on contextual pages without an explicit query, so the transformer's contribution must come purely from user history and page context---a harder regime in which to extract additional signal on top of an already-tuned production model.

\begin{table}[!ht]
  \centering
  \caption{Cross-surface comparison of the split scheme on desktop Search Ads and YAN web inventory. Runtime receptive field is $R = 100$ events; the offline encoder sees up to 8\,192 events on both surfaces. The split scheme retains a comparable fraction of the full-history ceiling on the two surfaces.}
  \label{tab:web-ablation}
  \begin{tabular}{lcc}
    \hline
    Configuration & Search Ads & YAN \\
                  & $\Delta_{\text{NLL}}$ (\%) & $\Delta_{\text{NLL}}$ (\%) \\
    \hline
    \rowcolor{tblref}
    Full model (8k, online ceiling) & +16.2 & +4.09 \\
    Split scheme (offline 8k + 100) & +11.7 & +3.26 \\
    \hline
    Retention vs.\ full ceiling     & 72\%  & 80\%  \\
    \hline
  \end{tabular}
\end{table}

The split scheme retains roughly $72\%$ of the full-history ceiling on desktop Search Ads ($11.7 / 16.2$) and $80\%$ on YAN web inventory ($3.26 / 4.09$). This consistency across surfaces, despite the structural difference noted above (query-conditioning on Search vs.\ contextual placement on YAN), suggests that the offline representation captures user-level patterns that transfer beyond a single surface rather than narrow surface-specific effects.

\subsection{RQ5: Does the magnitude of the gain track the amount of available history?}\label{sec:exp-puid}

The central premise of the paper is that long-range user histories carry predictive signal that shorter histories cannot reproduce. If this is true, then the same architecture should produce systematically larger gains for users whose histories are longer. We test this directly by stratifying the evaluation by user identifier type. User data is deduplicated either at the level of persistent user identifiers (PUIDs) for registered users, or device-level unique identifiers (UniqIDs) for anonymous users. PUIDs yield substantially richer interaction histories because they span sessions and devices, whereas device-level identifiers capture only shorter, single-device activity. Table~\ref{tab:puid} compares the resulting CTR-prediction quality across the two identifier types.

\begin{table}[!ht]
  \centering
  \caption{Impact of user identity type. PUIDs support longer histories and yield substantially larger gains.}
  \label{tab:puid}
  \begin{tabular}{lccc}
    \hline
    User type & Avg.\ history & $\Delta_{\text{NLL}}$ (\%) & \\
    \hline
    \rowcolor{tblref}
    PUID (registered)     & $\sim$1\,200 & +18.22 & \\
    UniqID (device-level) & $\sim$550    & +13.32 & \\
    \hline
  \end{tabular}
\end{table}

PUIDs carry roughly twice the average history length ($\sim$1\,200 vs.\ $\sim$550 events) and produce a correspondingly larger gain (+18.22\% vs.\ +13.32\%). This gap supports the central premise of the paper: richer long-range histories translate directly into better predictions. It also helps explain why the split architecture is especially valuable for users with persistent identities, for whom the amount of useful historical signal is largest.

\subsection{RQ6: Online A/B Test}\label{sec:exp-ab}

The online experiment provides the final validation of the proposed design. If the split architecture truly resolves the latency--quality trade-off, its offline gains should carry over to production without increasing serving cost. We therefore deployed the model in a production A/B test on two surfaces: Search Ads (desktop and mobile search listings) and YAN. The YAN deployment was carried out on \emph{YAN web inventory}, but online metrics are reported over the entire YAN network, since auction dynamics propagate the effect of a web-inventory ranking change to adjacent inventory; this gives a faithful picture of total business impact rather than a slice-localised number.

\begin{table}[!ht]
  \centering
  \caption{Online A/B test results on Search Ads and YAN. On YAN the model was deployed on YAN web inventory but the integral is measured over the full YAN network. All gains are statistically significant ($p < 0.05$).}
  \label{tab:ab-test}
  \begin{tabular}{lcc}
    \hline
    Metric & Search Ads & YAN \\
    \hline
    \multicolumn{3}{l}{\textit{Primary metric (acceptance)}} \\
    \rowcolor{tblbest}
    \quad Integral                & +2.77\%  & +2.10\% \\
    \hline
    \multicolumn{3}{l}{\textit{Clicks}} \\
    \quad Integral                & +2.87\%  & +2.59\% \\
    \hline
    \multicolumn{3}{l}{\textit{Revenue}} \\
    \rowcolor{tblhigh}
    \quad Integral                & +2.26\%  & +0.43\% \\
    \hline
  \end{tabular}
\end{table}

The split scheme improves the primary acceptance metric on both surfaces ($+2.77\%$ on Search Ads, $+2.10\%$ on YAN), with corresponding lifts in clicks and revenue (Table~\ref{tab:ab-test}). On Search Ads the lift is concentrated on mobile, where the production baseline relied less on long-history signal. On Search Ads, the click and revenue gains are roughly proportional. On YAN, click volume grows faster than revenue---a pattern that, taken alone, might suggest a shift toward lower-CPM impressions. The primary acceptance metric, however, is advertiser-side and tracks post-click conversions, and it improves in step ($+2.10\%$); this means the additional clicks are on average converting for advertisers. The widening click--revenue gap on YAN is therefore a desirable quality gain---the model selects clicks of higher advertiser value rather than diluting click quality.

No auxiliary metrics regressed during the experiment: all monitored indicators either improved or remained neutral. Importantly, serving latency did not increase. The offline model runs asynchronously on dedicated GPUs and never enters the request-time critical path. At request time, the runtime transformer is invoked in parallel with the other components of the existing CTR feature pipeline rather than being chained behind them, so its forward-pass cost overlaps with feature computation that the system already performs and is absorbed within the established latency budget. This is the main practical outcome of the paper: long-history transformer ranking can produce measurable online gains without increasing serving latency.

\subsection{Discussion}

The experiments support five practical conclusions:

\begin{enumerate}
\item \textbf{The split scheme is the key latency--quality mechanism.} Offloading long-history computation to an asynchronous offline model preserves between 72\% and 80\% of the full-model gain across the two evaluated surfaces while keeping runtime computation within real-time constraints.

\item \textbf{Sequence length is the main quality lever.} Extending the runtime receptive field improves quality more consistently than increasing model width, echoing findings from LLaTTE~\cite{xiong2026llatte} in a different production setting.

\item \textbf{The cached offline representation carries complementary long-range signal.} The large uplift from adding a single cached vector to the prediction head shows that the offline representation captures user patterns that cannot be reconstructed from the short runtime window alone.

\item \textbf{Training recipe matters as much as architecture.} Single-modality fine-tuning, organic click enrichment, and position debiasing together nearly double the gain, showing that data and optimization choices are first-order design decisions.

\item \textbf{Embedding staleness is manageable; model staleness is not.} Cached embeddings degrade slowly over days, enabling relatively inexpensive refresh policies, whereas stale model checkpoints lose quality much more rapidly and require regular retraining.
\end{enumerate}

\section{Conclusion}

This paper addresses a single practical question: how can long-history transformer models be used for CTR prediction in online advertising when full-history inference is too expensive for real-time serving? Our answer is a multi-stage architecture that separates heavy sequence computation from request-time inference. A large offline model processes the user's long-term history asynchronously and caches a fixed-size representation, while a lightweight runtime model combines that representation with fresh user activity under the production latency budget.

The experiments show that this split design resolves the main latency--quality trade-off effectively. Offline, it recovers between 72\% and 80\% of the quality of an impractical full-history runtime transformer across the two evaluated advertising surfaces (Search Ads and YAN). Operationally, the cached offline representation degrades slowly enough to support inexpensive refresh policies, while regular model retraining remains the main requirement for maintaining quality. In online A/B tests, the deployed system improves the primary metric by +2.77\% on Search Ads and by +2.1\% on YAN (deployed on web inventory, measured across the full network), with revenue gains of +2.26\% and +0.43\% respectively. Critically, serving latency did not increase, confirming that the split design fully decouples long-history modeling from request-time cost, and no auxiliary metrics degraded.

These results suggest that the main barrier to deploying trans\-former-based sequential models in ad ranking is not the usefulness of long user histories, but the cost of serving them naively. A split offline-online design makes those histories usable in production. While our implementation is specific to Yandex's serving infrastructure, the split-design pattern applies broadly to any latency-constrained ranking system with access to asynchronous computation.

Several directions remain open. Richer pre-training objectives---such as contrastive losses on content embeddings or auxiliary tasks predicting longer-term outcomes (post-click conversions, repeat-engagement signals)---could further improve the quality of the cached representation. Adaptive receptive-field allocation, where the online window size adjusts per user based on session recency, may improve the latency--quality trade-off for users with sparse recent activity. Finally, a more systematic study of scaling laws in this setting, analogous to the analysis in LLaTTE, could help quantify the interaction between model depth, sequence length, and feature richness and guide future capacity planning.


\bibliographystyle{ACM-Reference-Format}
\bibliography{bib}
\end{document}